\documentclass{rspublic}
\usepackage[dvips]{graphicx}

\begin{document}

\title[The circumstellar environment of rotating Wolf-Rayet Stars 
and the implications for GRB afterglows]{The circumstellar environment 
of rotating Wolf-Rayet Stars and the implications for GRB afterglows}

\author[J.J. Eldridge]{John J. Eldridge}

\affiliation{Astrophysics Research Centre, Queen's University Belfast, Belfast, BT7 1NN.}

\label{firstpage}

\maketitle

\begin{abstract}{Gamma-ray bursts; Massive stars; Stellar-wind bubbles.}
If Wolf-Rayet stars are the progenitors of Gamma-ray bursts (GRBs), they must rotate rapidly to produce the GRB. This rotation may effect their stellar-wind bubbles and possibly explain why so many GRB afterglows occur in a constant density medium.
\end{abstract}

\section{Introduction}

Wolf-Rayet (WR) stars (also know as helium stars) are the likely progenitors of GRBs. The more massive WR stars produce black holes at core-collapse and their radius is small enough for the relativistic jets to reach the stellar surface and produce the prompt GRB emission. WR stars have dense high-velocity winds that produce large stellar-wind bubbles through which the afterglow jet will propagate after the GRB event. Observations of afterglows agree with models of the stellar wind bubbles (e.g. Eldridge et al.(2006) and reference therein).

A problem remains that the free-wind region of the bubbles, where the density scales as $r^{-2}$, is always large and should be observed in every GRB afterglow. However for many GRB afterglows a constant density medium (CDM) has been inferred from afterglow observations. There is some uncertainty in estimating the circumburst environment but CDMs tend to be preferred. van Marle et al. (2006) have investigated how to move a CDM into closer proximity with the progenitor and found a number of possible effects such as stellar motion through the ISM.

However there is another possibility, the GRB progenitors must be rapidly rotating at the time of core collapse to ensure that the material around the forming black hole has enough angular momentum to produce an accretion disk. Stellar rotation can produce a strong effect on the stellar-wind bubble that has not been considered. Ignace et al. (1996) investigated the effect of rotation on stellar winds for various stellar types. They found that even moderate rotation of a WR star will effect the density of the wind at different latitudes on the stellar surface. This will effect the position of the wind termination shock by varying the ram pressure ($\rho v^{2}_{wind}$) with latitude. Their work is supported by observations of Wolf-Rayet binary CX Cephei. The light from the star is strongly polarised (approximately 4 percent) which was used to infer that the WR star is rapidly rotating producing a equator-to-pole wind density ratio of 5 (Villar-Sbaffi et al. 2006).

\section{Results}

Using the results of Ignace et al. (1996) we produced models of the distorted wind bubbles modifying the code used by Eldridge et al. (2006). Density profiles through two simulations are shown in figure 1. We see that if rotation reduces the polar wind density the distance to the CDM is reduced. Therefore in the rotating case the afterglow jet is more likely to inferred to be propagating through a CDM.

How close the CDM can move to the progenitor depends on how quickly the wind is accelerated and the rate of rotation. For a standard WR star wind the predictions in figure 1 are the result of a rotation rate of 45 percent of the stellar break-up velocity. We are currently working to determine how this effect changes for different WR stars. Such calculations are complicated as WR winds are optically thick so common assumptions of line driven winds cannot be applied. It is important to note that if the star is rotating close to break-up velocity then it may become highly distorted and the mass-loss geometry may become very different from that shown here (Owocki, Cranmer \& Gayley 1996).

\begin{figure}
  \includegraphics[angle=0,height=60mm]{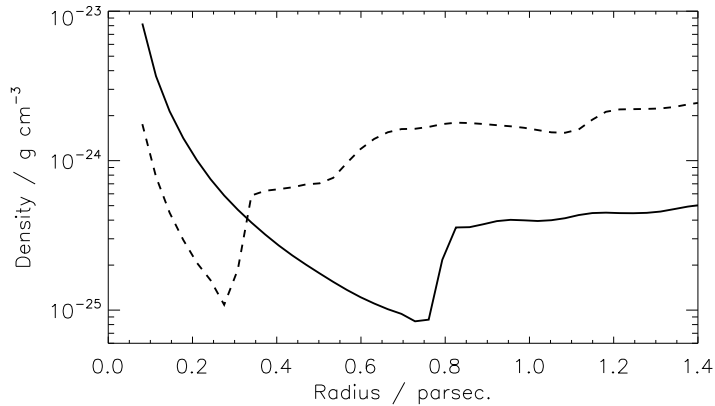}
  \caption{Density profiles through two stellar-wind bubbles along the assumed rotation axis. The solid line is the non-rotating star. The dashed line is the rotating star, with a equator-to-pole density ratio of 5. The WR star's mass is 11M$_{\odot}$ with $V_{wind} = 1800 \, km \, s^{-1}$ and $\dot{M} = 1.2 \times 10^{-5}\, M_{\odot} \, yr^{-1}$ and the initial ISM density was $10^{3} cm^{-3}$.}
\end{figure}

\begin{acknowledgements}
JJE thanks Allard-Jan van Marle, Norbert Langer, Jorick Vink and Paul Crowther for useful discussion and advice and Robert Mochkovitch for starting him on this project.
\end{acknowledgements}

\end{document}